\documentclass[apj]{emulateapj}

\shorttitle{Stellar Content \& Recent SFH of IC1613}
\shortauthors{Bernard et al.}

\begin{document}

\title{Stellar Content and Recent Star Formation History of \\ the Local Group
 Dwarf Irregular Galaxy IC1613\footnotemark[1]}

\author{Edouard J. Bernard, Antonio Aparicio\altaffilmark{2}, Carme Gallart,
        Carmen P. Padilla-Torres, Maurizio Panniello$\dagger$}
\affil{Instituto de Astrof\'{i}sica de Canarias, E-38205 La Laguna, 
       Tenerife, Spain}
\email{ebernard@iac.es, antapaj@iac.es, carme@iac.es, cppt@iac.es.}

\footnotetext[1]{Based on observations made with the Isaac Newton Telescope
 operated on the island of La Palma by the Isaac Newton Group in the Spanish
 Observatorio del Roque de los Muchachos of the Instituto de Astrof\'{i}sica
 de Canarias.}
\altaffiltext{2}{Departamento de Astrof\'{i}sica, Universidad de La Laguna,
                 E-38200 La Laguna, Tenerife, Spain.}
\altaffiltext{$\dagger$}{Deceased.}

\begin{abstract}

    We present resolved-star $VI$ photometry of the Local Group dwarf
    irregular galaxy IC1613 reaching I\,$\sim$\,23.5, obtained with the
    wide-field camera at the 2.5\,m Isaac Newton Telescope.
    A fit to the stellar density distribution shows an exponential profile
    of scale length $2\farcm9 \pm 0.1$ and gives a central surface
    brightness $\mu_{V,0} = 22.7 \pm 0.6$.
    The significant number of red giant branch (RGB) stars present in the
    outer part of our images ($r > 16\farcm5$) indicates that the galaxy
    is actually more extended than previously estimated.
    A comparison of the color-magnitude diagrams (CMDs) as a function of
    galactocentric distance shows a clear gradient in the age of its
    population, the scale length increasing with age, while we find no
    evidence of a metallicity gradient from the width of the RGB.
    We present quantitative results of the recent star formation history
    from a synthetic CMD analysis using IAC-STAR. We find a mean star formation
    rate of $(1.6 \pm 0.8) \times 10^{-3} $~M$_\Sun~$yr$^{-1}~$kpc$^{-2}$ in
    the central $r \la 2\farcm5$ for the last 300~Myr.

\end{abstract}


\section{Introduction}

\setcounter{footnote}{2}

    IC1613 is a typical dwarf irregular galaxy concerning its
    luminosity, metallicity and star formation rate (SFR). In fact, it 
    serves as the prototype for the DDO type Ir~V.
    It is a low surface-brightness galaxy with a moderate luminosity
    \citep[$M_V=-15.3$,][]{ber00} located $730 \pm 20$ kpc from our
    Sun. This distance corresponds to a distance modulus
    $(m-M)_0=24.31 \pm 0.06$ \citep{dol01} and a scale of 3.54 pc per
    arcsecond. The recent calculation of \citet{pie06} from near-IR
    photometry of cepheids gives $(m-M)_0=24.291 \pm 0.035$, for which
    the authors claim a total uncertainty of less than 3\%.
    While the spatial extent estimate of \citet{abl72} gives an optical
    size of $16\arcmin \times 20\arcmin$, the observation of carbon stars
    up to 15$\arcmin$ from the center by \citet{alb00} indicate that it
    is virtually twice that size.
    Its high galactic latitude, in the southern hemisphere, confers it
    low extinction and color excess. Here we adopt a reddening of
    $E(B-V)=0.02 \pm 0.02$ from \citet{col99}.

    The H\,\textsc{II} region metallicity of IC1613 was measured by 
    \citet[see \citealt{ski89}]{tal80} 
    to be 12\,+\,log(O/H)~$=7.86$ from spectrophotometric observations
    of the [O\,\textsc{III}]$\lambda4363$ line, while \citet{lee03}
    found 12\,+\,log(O/H)~$=7.62$.
    This corresponds to [Fe/H]~$=-0.8$ and [Fe/H]~$=-1.07$ dex
    respectively \citep[hereinafter S03]{ski03}.
    The mean color $(V-I)_{-3.5}$ of the RGB at $M_I$~=~-3.5 gives
    [Fe/H] = -1.3 for the old and intermediate-age population \citep{lee93}.
    This low overall metallicity and the high gas content \citep{hof96} 
    suggest a primitive state in its evolution.

    \defcitealias{ski03}{S03}
    \defcitealias{col99}{Cole et al. 1999}
    \defcitealias{dol01}{Dolphin et al. 2001} 
    
    The star formation history (SFH) has been studied quite extensively by
    \citetalias{ski03} for an HST/WFPC2 field located $7\farcm4$ kpc southwest
    from the center. Their conclusion is a relatively constant SFR over a long
    period, with the oldest population being more than 10~Gyr old.
    They also summarize the results about structure and stellar content of
    the whole galaxy from the literature. More recently, \citet{bor04}
    analyzed 60 OB associations, apparently correlated with H\,\textsc{II}
    regions studied kinematically by \citet[see also \citealt{sil06}]{loz03}.
    \citet{mag05} announced the detection of two candidate
    planetary nebulae.
    
    In this paper we describe a wide-field survey of IC1613: an overview of
    the observations and data reduction is presented in section
    \ref{datared}, and the resulting color-magnitude diagram (CMD) is 
    described in \S \ref{descCMD}.
    In \S \ref{morpho} we examine the extent and morphology of the galaxy,
    as well as the spatial structure of the different populations in the CMDs. 
    An analysis of the recent SFH at different galactocentric radii is given
    in \S \ref{recentstarfo}.
    Finally, in \S \ref{concl} we summarize the results and present
    our conclusions.

\section{Observations \& Data Reduction}\label{datared}

    Observations of IC1613 in Harris $V$ and Sloan Gunn $i'$ were conducted on
    5 nights between November 1999 and September 2000 using the Wide Field
    Camera (WFC) at the 2.5 m Isaac Newton Telescope (INT) of the
    Observatorio del Roque de los Muchachos.
    The WFC is a mosaic camera made up of four $2048 \times 4096$ CCDs, with
    a pixel size of $0.33 \arcsec$. The total field of view is about $34
    \arcmin \times 34 \arcmin$, covering most of the galaxy. The total
    integration times in $V$ and $i'$ were 3660 and 1830 seconds, respectively.
    A detailed observing log is presented in Table~\ref{obs}.

\begin{deluxetable}{ccccc}
\tablewidth{0pt}
\tablecaption{Journal of Observations. \label{obs}}
\tablehead{
\colhead{UTC Date} & \colhead{Time (UT)} & \colhead{Filter} &
\colhead{Exp. Time (s)} & \colhead{Air Mass}}
\startdata
1999 Nov 06 & 22:23 & V &  60  & 1.14 \\
1999 Nov 06 & 22:32 & V & 1200 & 1.13 \\
1999 Nov 06 & 22:53 & V & 1200 & 1.12 \\
1999 Nov 06 & 23:14 & V & 1200 & 1.12 \\
2000 Aug 10 & 04:38 & I &  30  & 1.12 \\
2000 Aug 10 & 04:41 & I & 600  & 1.12 \\
2000 Aug 10 & 04:54 & I & 600  & 1.12 \\
2000 Aug 10 & 05:05 & I & 600  & 1.12 \\
\enddata
\end{deluxetable}

    Overscan trimming, bias subtraction and flat-field corrections were
    performed using the standard routines in IRAF\footnote[3]{IRAF is
    distributed by the National Optical Astronomy Observatory,
    which is operated by the Association of Universities for Research
    in Astronomy, Inc., under cooperative agreement with the National
    Science Foundation.}. The $i'$ images have also
    been corrected for fringing effects. The DAOPHOT-II/ALLSTAR and ALLFRAME
    programs \citep{ste87,ste94} were then used to obtain the instrumental
    photometry of the resolved stars from the four individual images in each
    band and for each chip. The $\sim$200 stars used to model the
    point-spread functions (PSFs) were carefully selected to cover the whole
    field of view and sample the spatial variations of the PSF. The input
    list of stars for ALLFRAME was created with DAOMASTER
    from the ALLSTAR photometry files of the individual images. This list
    contains all the stars that were detected on at least one image.
    The stars with good photometry were selected among the detected objects
    using ALLFRAME's fitting parameters $\sigma$, $\chi ^2$ and $SHARP$. Only
    those $\sim$30000 objects with very good photometry, i.e.,
    with $\sigma \leq 0.15$, $\chi ^2 \leq 1.1$
    and $-0.3 \leq SHARP \leq 0.3$, were kept.

\begin{figure}[b]
\epsscale{0.8} 
\epsscale{1.1} 
\plotone{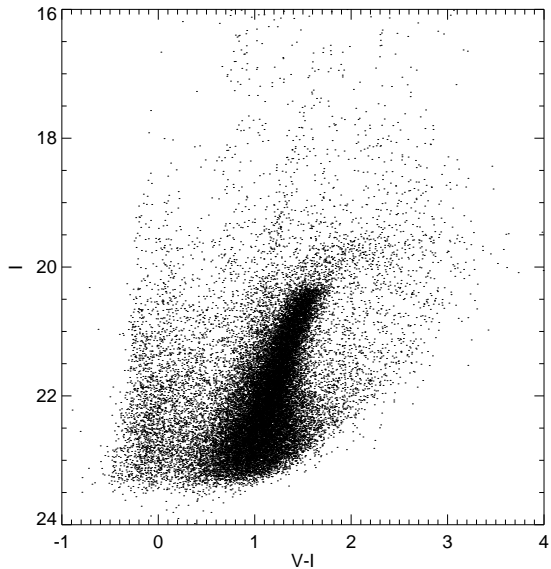}
\figcaption{Color-magnitude diagram of IC1613. All the stars described
 in section \ref{datared} are plotted.
\label{CMD}}
\end{figure}

\begin{figure}
\epsscale{0.8} 
\epsscale{1.1} 
\plotone{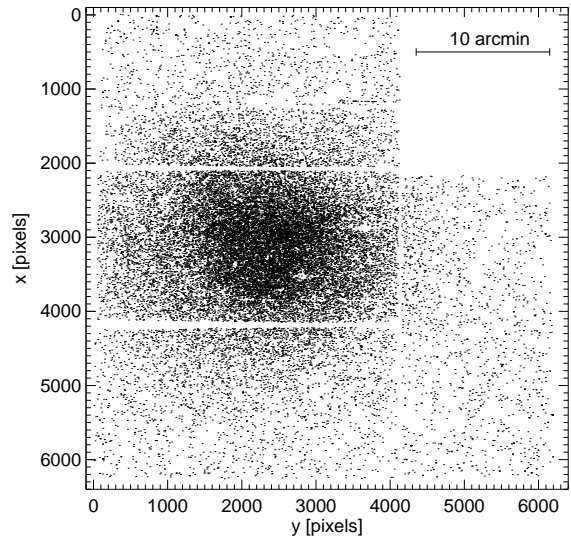}
\figcaption{Distribution of resolved stars in IC1613.
 North is up and East to the left.
\label{star_distrib}}
\end{figure}

    Our photometry of chips 1, 3 and 4 of the WFC\footnote[4]{See the WFC
    User Notes at
    http://www.ing.iac.es/Astronomy/instruments/wfc/wfc\_notes\_apr98.html.} 
    was calibrated to standard magnitudes using OGLE's photometry of the same
    field, kindly provided by Dr.~A. Udalsky (private communication).
    Several hundreds of stars were used for each chip, giving dispersions
    of the fits at the centers of mass of the point distributions of about
    0.001.
    Calibration of chip 2, which was outside OGLE's field, was more laborious.
    In $V$, one standard star field observed during the IC1613 INT run
    was used to determine the transformations between chip 2 and chips 1 and 4.
    The dispersion of these transformations is $\sim$0.005.
    Then the transformation for the latter chips based on OGLE's photometry was
    used, bringing the chip 2 $V$ photometry into the standard system.
    In $I$, chip 2 was calibrated differentially with respect to
    chip 4 using overlapping images obtained on the IAC80 telescope
    at Iza\~{n}a, Tenerife, Spain, during a photometrical observing run.
    The dispersion at the center of mass is of the order of 0.01.
    Hence, the total error in our photometry is that given by \citet{uda01},
    i.e., up to 0.02 for both $V$ and $I$ bands.
    Figure~\ref{CMD} shows our final ($V$-$I$, $I$) CMD\footnote[5]{The
    photometric data are available from the first author upon request.}.
    The spatial distribution of these stars is presented in
    Fig.~\ref{star_distrib}.

    The errors given by ALLFRAME are the residuals of the PSF-fitting
    procedure, so they should be considered internal errors. Signal-to-noise
    limitations, stellar crowding, blending and starloss, which we can refer to
    as observational effects, are important error sources and significantly
    modify the CMD shape and stellar density distribution \citep{apa95}.
    To estimate the observational effects and the completeness of our
    photometry, we resorted to artificial stars tests. 
    See \citet{apa95} for a detailed description of the procedure
    and the effects of crowding.
    Basically, a large number of artificial stars covering the same range in
    color and magnitude as the observed stars is added to the images using the
    corresponding PSF. These were placed on the images following a triangular
    grid in order to avoid crowding between artificial stars
    themselves and to optimally sample the chip fields. The photometry is then
    repeated in the exact same way as was done originally. A comparison of the
    input and output artificial-star list gives information about the 
    completeness as a function of magnitude and galactocentric distance.

\begin{figure}
\epsscale{1.0} 
\epsscale{1.25} 
\plotone{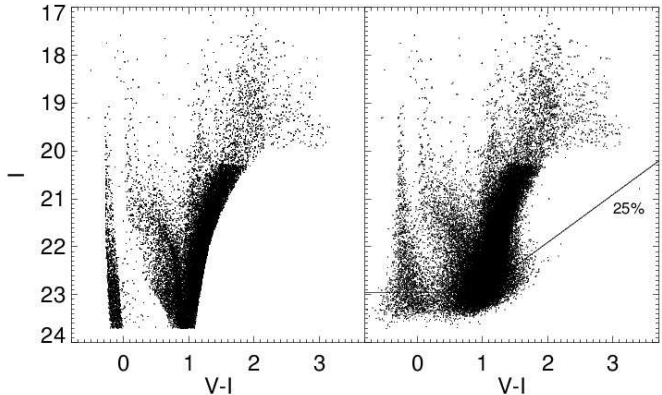}
\figcaption{{\it Left:} Synthetic CMD produced by IAC-STAR showing the
 153260 stars injected in chip 4 ({\it see text for details}).
 {\it Right:} CMD of the $\sim$40000 {\it artificial} stars recovered both
 in V and I and considered to have good photometry as described in
 section 2. The line shows the 25\% completeness limit as determined from the
 artificial star tests.
\label{inj_rec_stars}}
\end{figure}

\begin{figure}
\epsscale{0.8} 
\epsscale{1.1} 
\plotone{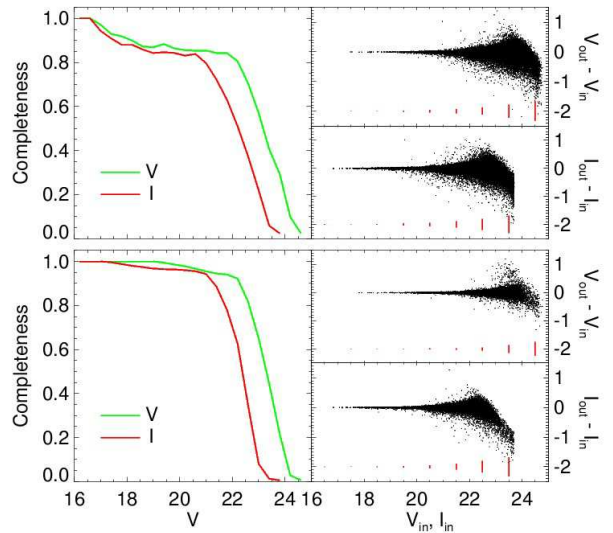}
\figcaption{Completeness ({\it left}) and errors in recovered magnitudes 
 ({\it right}) for a central field (chip~4, {\it top panel}), and an outer
 field (chip~1, {\it bottom panel}). The errorbars in the right panels show
 the dispersion per magnitude bin.
\label{err}}
\end{figure}

    As the artificial star sample, we took a synthetic CMD produced by
    IAC-STAR\footnote[6]{The code, available for free use, can be executed
    at the IAC-STAR website http://iac-star.iac.es/.} \citep{apa04}
    using the stellar evolution library of \citet{ber94} and the
    bolometric corrections of \citet{cas03}. The SFR was chosen
    constant between 13~Gyr ago and now, while the metallicity range increases
    from 0.0008 to 0.002 at $t=0$, to 0.0008 to 0.006 at the present time.
    These ranges were chosen wide enough so that they include the actual
    metallicity of IC1613. In total, 76630 artificial stars
    were added in 5 runs in each external chip, and twice as many in
    the central chip (\#4) where crowding is more important. 
    Figure~\ref{inj_rec_stars} presents the injected and recovered CMDs for
    chip 4. The broken line indicates the 25\% completeness limit, averaged
    over the galactocentric radius range, obtained from the ratio of recovered
    to injected stars as a function of magnitude. Figure~\ref{err} shows the
    completeness and error in recovered magnitude as a function of input
    magnitude for an inner and an outer field. It is important noticing that,
    even though the recovered artificial stars were filtered using the same
    values of $\sigma$, $\chi ^2$ and $SHARP$ as the real stars, {\it external}
    errors can be as large as $\sim$1~magnitude at the faint limit.

\section{The Color-Magnitude Diagram}\label{descCMD}

    The most evident feature of our CMD, Fig.~\ref{CMD}, is the red giant 
    branch (RGB), composed of low mass stars 
    \citep[M~$\la 1.8-2.0M_{\Sun}$;][]{chi92} older than about 1~Gyr and
    burning hydrogen in a thin layer around an inert helium core. 
    The age-metallicity degeneracy and the presence of the fainter extension
    of the asymptotic giant branch (AGB) in the same region of the CMD makes it
    difficult to get detailed information about the stars populating it.
    However, it is possible to place rough limits in time for given
    metallicities using theoretical isochrones.
    A more detailed determination of the chemical enrichment law (CEL) using
    this method, as well as its limitations, is presented in section
    \ref{recentstarfoMet}. \\
    The presence of a well populated main sequence (MS) blueward of $V-I=0$ and
    up to $I \sim 18.5$ indicates very recent star formation ($\la$10~Myr). \\
    The stars with $0 \leq (V-I) \leq 0.6$ and $I \leq 22$ are most likely blue
    loop (BL) stars, highlighting the blue edge of the core He-burning loop,
    while the red edge (i.e., the red supergiant branch, RSG), although
    contaminated by foreground stars, is well defined from $(V-I) \sim 1.0$ to
    $(V-I) \sim 1.8$ and up to about $I \sim 16.5$.
    These are young and intermediate- to high-mass stars, so
    they are among the most metal-rich stars in IC1613. However, this
    is a poorly understood phase in stellar evolution and the theoretical
    models still contain large uncertainties owing to the importance of
    processes such as mass-loss, overshooting and rotation in very massive
    stars \citep[see e.g.,][]{mae01}. The BL and RSG actually reach the red
    clump down to $I=23.76$ \citep{col99}, but the dispersion at the faint
    end of our CMD makes it impossible to distinguish them from the RGB stars
    below $I=22$.\\
    The final extension of the AGB, or red-tail \citep{gal94}, extends
    horizontally redward from the RGB tip at $I \sim 20$. AGBs are shell H- and
    He-burning stars, of low and intermediate mass and age over about 0.1~Gyr. 
    Their relatively large number indicates a possibly important
    intermediate-age population with relatively high metallicity, which would
    be compatible with the enhanced SFR between 3 to 6~Gyrs ago found by
    \citetalias{ski03}.

\begin{figure}
\epsscale{0.48} 
\epsscale{1.0} 
\plotone{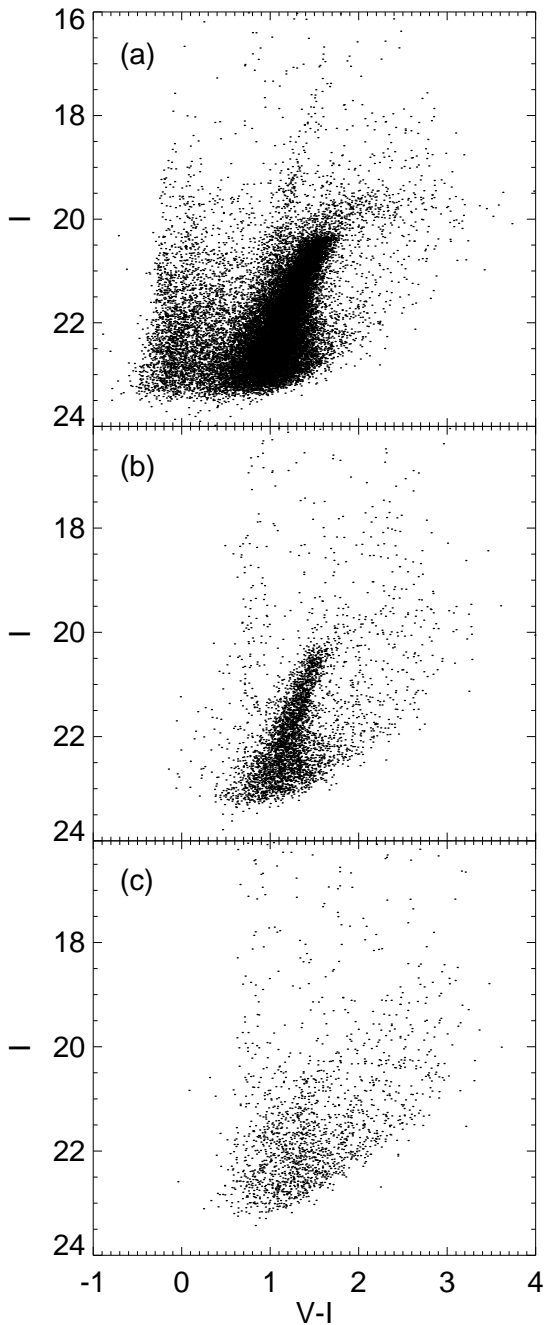}
\figcaption{Color-magnitude diagram of IC1613 for different galactocentric
 distances. The radius ranges are
  {\it (a)}~$r \leq 10\arcmin$,
  {\it (b)}~$10\arcmin < r \leq 16\farcm5$,
  {\it (c)}~$r > 16\farcm5$.
\label{3CMDs}}
\end{figure}

    Because of the wide field of view, the diagram also contains a relatively
    large number of foreground stars.
    Most stars with $0.5 \le V-I \le 1.2$ and $I \le 20$ are
    probably Galactic dwarfs since their number is the same for the middle
    and bottom panels in Fig.~\ref{3CMDs} after correcting for the
    difference in area. More generally, the foreground contamination is
    rather important for $V-I \ga 0.6$.

    Although old stars ($\ga$10~Gyr) are very likely present in the RGB,
    our CMD is not deep enough to detail the old, low metallicity population
    as it would if fainter MS stars were observed. The existence of a {\it bona 
    fide} old population in IC1613 is known thanks to the presence of RR Lyrae
    stars \citep{sah92,dol01} and CMDs from Hubble Space Telescope observations
    showing core helium burning, horizontal branch stars
    \citepalias{col99,dol01,ski03} and the oldest MS turnoffs \citep{gal07}.

\section{Morphology, Spatial Extent \& Distribution of Stellar Populations}\label{morpho}

    To characterize the morphology of IC1613, we plotted the RGB star
    distribution obtained from our photometry, and convolved the resulting
    map with a Gaussian of $\sigma=50 \arcsec$ using IRAF's GAUSS from the
    IMFILTER package. The result of this process is a smooth map of the
    stellar density highlighting the morphology of the galaxy. Fitting
    ellipses to the isodensity contours was done with IRAF's ISOPHOTE routine.

    From the best fitted ellipses, where the crowding is low but
    the star number sufficient, we find a position angle of 80$\degr$ and an
    eccentricity $\epsilon=1-b/a=$ 0.15, in good agreement with the
    values given by \citet[$PA=81 \degr$, $\epsilon=0.18$;][]{abl72}.
    Following the shape of the isopleths, we divided the galaxy into concentric
    ellipses at small radii ($\la$10$\farcm5$) and circles at larger ones
    that we used for radial star counts and stellar populations gradient
    analysis. Their semi-major axis increases in steps of 100 pixels,
    corresponding to $33 \arcsec$.

\begin{figure}
\epsscale{0.8} 
\epsscale{1.1} 
\plotone{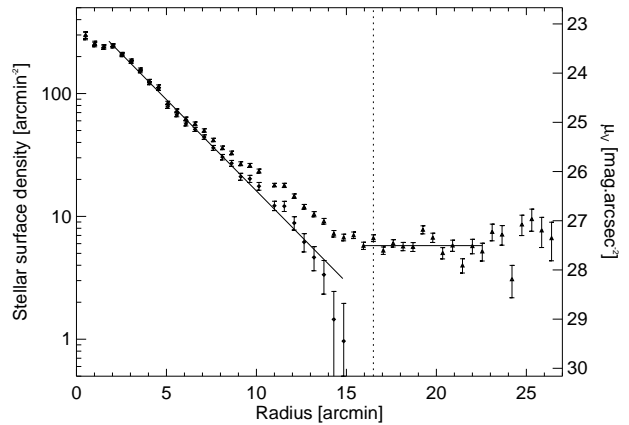}
\figcaption{Number density of stars as a function of galactocentric distance
 before ({\it triangles}) and after ({\it diamonds}) background subtraction.
 The right-hand vertical axis gives a rough estimate of the surface magnitude, 
 calibrated as described in text.
 The horizontal line is the weighted mean density of ellipses 29 to 41, which
 has been adopted as the background level. The exponential fit has a scale
 length of $2\farcm9\pm 0.1$. The vertical dotted line shows where the
 foreground stars start to dominate, as obtained from the CMDs (see section
 \ref{morpho}).
\label{nb_density}}
\end{figure}

    Figure~\ref{nb_density} shows the radial profile of the galaxy constructed
    from the number of stars in each ellipse after correcting for completeness.
    The correction was obtained from the ratio of the number of recovered
    to injected stars in the artificial star test for each annular region.
    The area of the ellipses has been calculated via Monte-Carlo sampling,
    carefully taking into account the gaps between the chips as well as the
    regions around saturated stars when calculating the effective surface. 
    An approximate surface brightness scale, shown on the right-hand side of
    Fig.~\ref{nb_density}, was calculated from the
    stellar density in each ellipse. The calibration was calculated by
    comparing the star number with the total, sky subtracted flux
    for each ellipse then averaged over the radius range for consistency.
    An exponential least square fit to this curve between $2\arcmin$ and 
    $15\arcmin$ from the center gives a scale length of
    $2\farcm9 \pm 0.1$ ($620\pm20$ pc) and central surface brightness
    $\mu_{V,0} = 22.7 \pm 0.6$. The former value is slightly smaller
    than the $760\pm50$ pc ($\sim$3$\farcm5\pm0.2$) given by \citet{hod91}.
    However, a similar larger length is obtained when fitting only the inner
    $\sim$7$\arcmin$, as was done by the authors because of the limited depth
    of their photometry, and omitting the crowding correction.
    The profile seems to get steeper at $r \sim 12\arcmin$, but this change
    of slope could be an artifact of the background subtraction and small
    number statistics.


\begin{figure}
\epsscale{0.56} 
\epsscale{1.09} 
\plotone{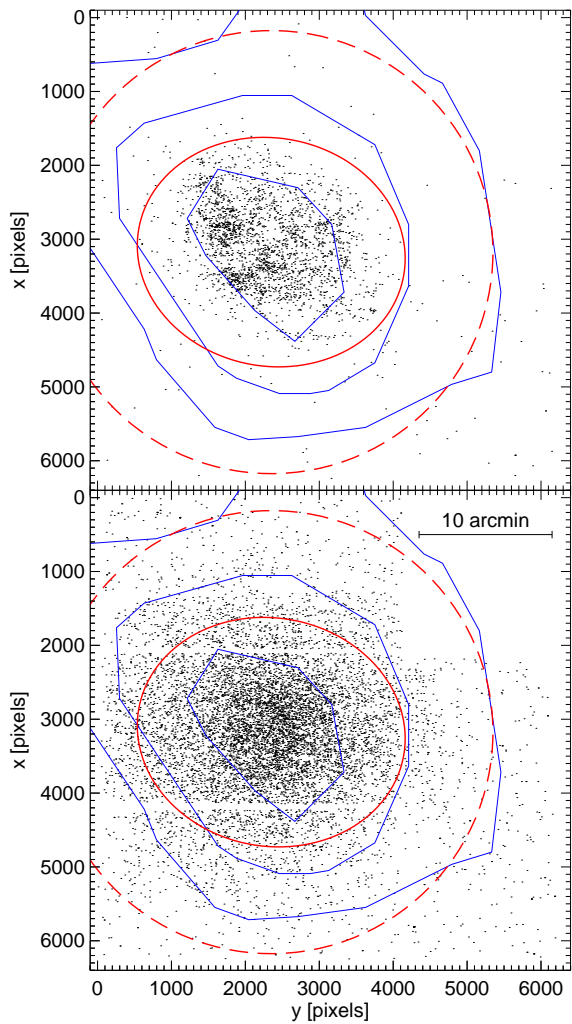}
\figcaption{Spatial distribution of young (MS+BL+RSG; {\it top}) and older
 (AGB+RGB; {\it bottom}) stars.
 The red ellipse shows the extent of the MS stars (r=10\arcmin).
 The red dashed circle indicates where the foreground stars
 start to dominate (r=16\farcm5).
 The neutral hydrogen emission contours (0.2, 6.2 and 14.2 Jy km s$^{-1}$)
 from \citet{hof96} have been overplotted in blue ({\it see text for details}).
\label{star_distPop}}
\end{figure}

    The large field of view covered by the WFC permits to study the spatial
    gradients of the stellar population across the galaxy. Ideally, that would
    give us hints on its formation and evolution. However, in the case of
    shallow CMDs, the spatial variations of the morphology of the CMD only
    reflect accurately differences in the SFH over the last several hundred
    million years.
    Such variations have been observed in all the known dwarf irregular
    galaxies through CMD morphology (e.g., WLM: \citealt{min96}; NGC 6822:
    \citealt{bat06}; Leo A: \citealt{van04}; Phoenix: \citealt{mar99}) or
    distribution of the variable star populations (e.g., in Phoenix:
    \citealt{gal04}; in Leo I: \citealt{bal04}).

    Figure~\ref{3CMDs} shows that it is also the case for IC1613 and
    confirms the difference in relative number of young and old stars found
    by \citetalias{ski03} between their outer HST field and a central field 
    studied earlier by \citet{col99}. We divided the galaxy into three regions
    following the morphology of the CMD of each individual ellipse defined
    above: the inner part of the galaxy where stars younger than about
    500~Myr are present ($r \leq 10\farcm1$, Fig.~\ref{3CMDs}a), the region
    at intermediate distance with no young star but still a well defined RGB
    ($10\farcm1 < r \leq 16\farcm5$, Fig.~\ref{3CMDs}b), and the outermost
    part of our observed field, dominated by foreground stars 
    ($r > 16\farcm5$, Fig.~\ref{3CMDs}c). The corresponding spatial limits,
    as well as the H\,\textsc{I} contours from \citet{hof96}, are displayed
    over the stellar distribution of the young and old populations in
    Fig.~\ref{star_distPop}. 
    Note that the H\,\textsc{I} contours were shifted by $-0\farcm6$ (+100
    pixels in y) and $2\farcm9$ ($-$500 pixels in x) in right ascension and
    declination, respectively. This corresponds to the offset between the
    astrometry of \citet{lak89} and \citet{hof96}, and adopting the former
    for the H\,\textsc{I} contours correctly places the H\,\textsc{II} regions
    of \citet{hod90} on top of the star forming regions, and fits our star
    distribution better.
    Although the RGB of Fig.~\ref{3CMDs}c is not clearly defined, a substantial
    fraction of the displayed stars probably belong to IC1613. This
    corroborates the results of \citet{alb00}, who found carbon stars 
    extending up to $15\arcmin$ from the center of the galaxy, and shows
    that IC1613 is actually more extended than previously thought.

\begin{figure}
\epsscale{0.8} 
\epsscale{1.15} 
\plotone{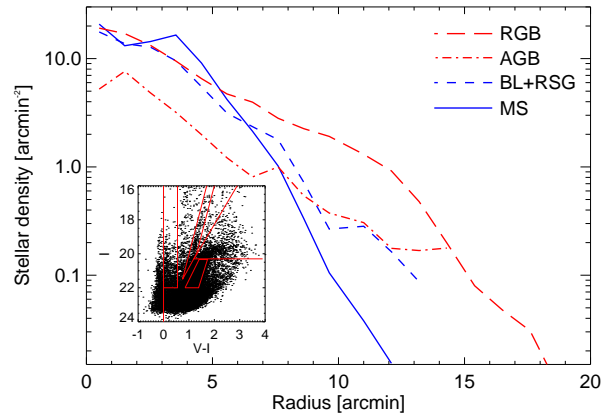}
\figcaption{Stellar densities of the different populations of resolved
 stars after correction for completeness and background contamination. 
 The stars selected for each population are shown in the inset. 
 The RGB profile was divided by four to fit on the graph. 
 The errorbars have been omitted for clarity. 
 Note the difference in the scale length of the older (RGB, AGB) and younger 
 (MS, BL+RSG) populations ({\it see text for details}).
\label{pop_densities}}
\end{figure}

    To give a quantitative measure of the gradient in the age composition of
    the stars in IC1613, the stellar surface
    density for different populations of resolved stars is presented in
    Fig.~\ref{pop_densities}. The age gradient is clearly
    visible: while the density of the older stars follows the expected
    exponential decrease from the central region, that of the young population
    peaks at a radius of $\sim$3$\arcmin$ and vanishes rapidly as the radius
    increases. This results in the scale length of the young population being
    much smaller than that of the older stars: a fit to the profiles
    between $3\farcm5$ and $12\farcm1$ from the center of IC1613 gives scale
    lengths of $1\farcm19 \pm 0.04$, $2\farcm0 \pm 0.1$, $3\farcm2 \pm 0.2$
    and $3\farcm8 \pm 0.1$ for the MS, BL+RSG, AGB and RGB, respectively.
    The off-centered peak in the distribution of the young population, visible
    at $r\sim3\arcmin$--$3\farcm5$ in Fig.~\ref{pop_densities}, is due
    to the fact that the star forming regions are distributed in a somewhat
    circular pattern at this distance from the center \citep{hod90}, where
    \citet{sil06} observed a higher H \textsc{I} column-density.


\begin{figure}
\epsscale{0.8} 
\epsscale{1.1} 
\plotone{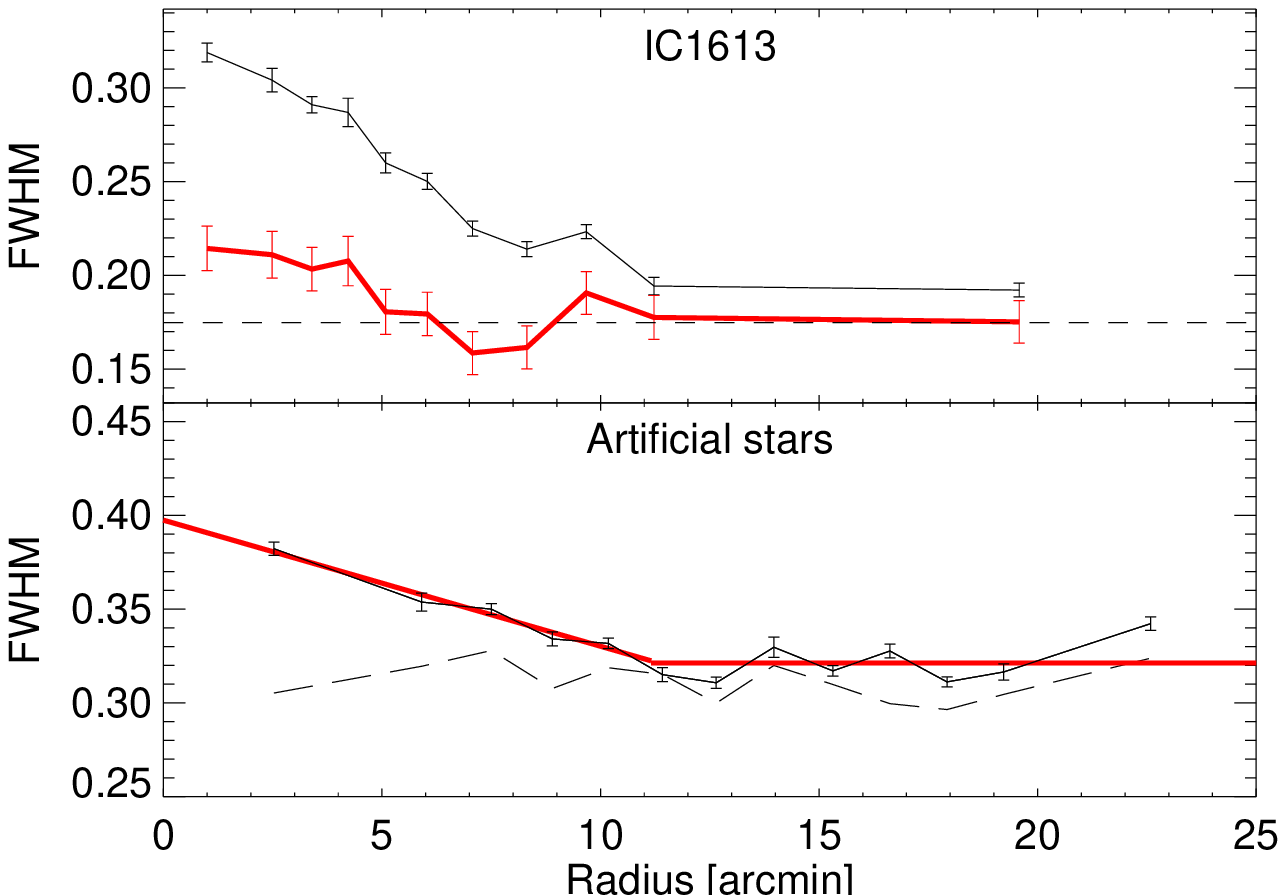}\\
\plotone{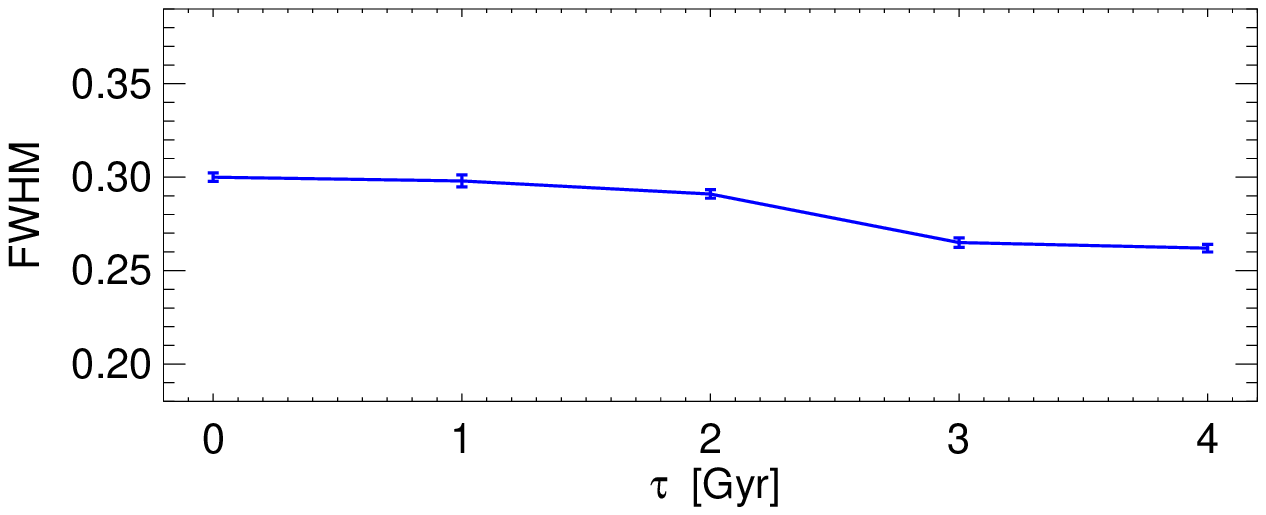}
\figcaption{{\it Top:} Width of the RGB of IC1613 at $M_I=-3.5$ as a function
 of galactocentric distance, before ({\it thin line}) and after ({\it thick
 red line}) correction for the effects of crowding. The dashed line indicates
 the mean value of the corrected width beyond $\sim$$5 \arcmin$. 
 {\it Middle:} Width of the injected ({\it long-dashed line}) and recovered
 ({\it full thin line}) RGB of the artificial stars tests. 
 The thick red line is a fit to the recovered width.
 {\it Bottom:} Width of an artificial RGB as a function of the age of the
 youngest stars in the CMD. The small change between $\tau=0$ and $1$~Gyr is
 probably due to the presence of stars from the RSG sequence. In all cases the
 errorbars correspond to the dispersion of twenty solutions.
 The vertical scale is the same on all three panels.
\label{RGBwidth}}
\end{figure}

    Because of the large dependence of the RGB color on metallicity, the
    position of the RGB in a CMD can be used to estimate the metallicity of
    the stellar system using empirical relations \citep[e.g.,][]{dac90}. For
    the same reason, the width of the RGB is often considered to be an
    indication of the metallicity dispersion.
    On the CMDs presented in Fig.~\ref{3CMDs}, it appears that the RGB in
    the central region is wider than that at larger radii and suggests the
    existence of a metallicity gradient.
    However, a RGB width gradient may also reflect an observational-effects
    gradient. To check this possibility, the following procedure was used,
    including measurements of real- and artificial-star RGB width and error
    estimates. We measured the width of the RGB at $M_I=-3.5$ as a function of
    galactocentric distance for the real stars as follows:
    for each radius interval, chosen so that the RGB contains about 800 stars,
    a subsample of 450 RGB stars was selected at random.
    The resulting RGB was sliced in intervals of 0.25 magnitude between
    $20.5 \leq I \leq 22$, and the width of the RGB in each magnitude range
    was obtained from the full-width at half-maximum (FWHM) of a Gaussian fit
    to the color function (CF). The width at $M_I=-3.5$ was then interpolated
    from a linear fit to the FWHM versus magnitude plot. We repeated this
    operation twenty times using a different subsample of RGB stars each time,
    and used the dispersion of these solutions as the errorbars. The same
    process was followed for the synthetic CMD used in the artificial stars
    tests, except that the radius bins
    and random subsamples contained 2500 and 1500 stars, respectively.
    The resulting plot is presented in Fig.~\ref{RGBwidth} for IC1613
    ({\it top}) and the artificial stars ({\it middle panel}). 
    
    It shows that the RGB recovered in the artificial star tests is wider
    in the center of the galaxy. In addition to the expected FWHM
    widening due to signal-to-noise limitations, crowding in the central part
    further disperses the stars on the CMD. Crowding is important and affects
    the width of the RGB up to $r\sim 10 \arcmin$. 
    To correct the observed RGB for the effects of crowding, we calculated
    the broadening parameter as a function of radius by subtracting in 
    quadrature the injected width from the recovered one. The corrected width
    was then obtained by subtracting in quadrature the broadening parameter
    from the observed RGB width, and is shown as the thick red line in the
    upper panel of Fig.~\ref{RGBwidth}.
    The FWHM of the corrected RGB still presents a significant
    variation across the central $\sim$6$\arcmin$.
    A higher metallicity dispersion in the central region could be responsible
    for this residual width excess. However, the lower panel of
    Fig.~\ref{RGBwidth}, showing the width of artificial RGBs for which the
    age of the youngest stars is progressively older, indicates that this
    excess is consistent with the presence of stars between 1 and 3--4 Gyr old
    in the center.
    Thus, there is no strong evidence of a metallicity gradient across the
    galaxy but its presence cannot be ruled out with the present set of data.

\section{Recent Star Formation History}\label{recentstarfo}

\subsection{The Method}\label{recentstarfoMet}

    The reconstruction of the recent SFH has been performed in a way similar
    to that described in Hidalgo \& Aparicio (2007, in preparation) or
    \citet{gal99}. The main difference here resides in the fact that our CMD
    is not sufficiently deep to contain discriminating information about
    intermediate-age and old stars. We limited our study to the recent SFH,
    i.e., the last $\sim$300~Myr, based on the MS, BL and RSG populations. 
    It thus limited the determination of the SFH to the inner 
    r~$\la 10\arcmin$ since these populations are absent beyond this radius.

\begin{figure}
\epsscale{0.8} 
\epsscale{1.2} 
\plotone{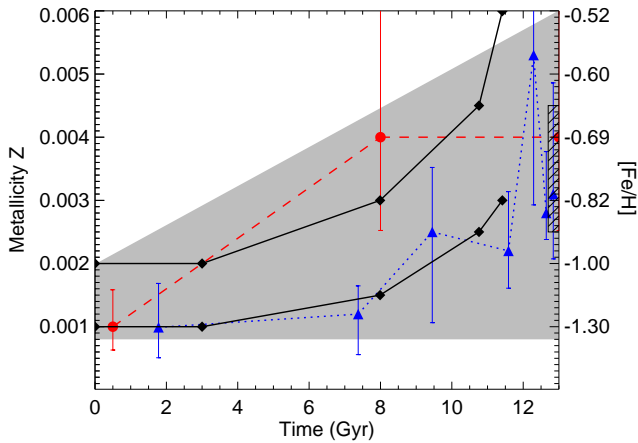}
\figcaption{CEL of IC1613. The diamonds indicate, for a given age, the
 minimum and maximum values found by fitting theoretical isochrones from the
 library of Padua \citep{gir02} to the upper RGB. For reference, the CELs
 obtained via the `Tolstoy' and `Dolphin methods' from \citetalias{ski03} are
 shown as red circles and blue triangles, respectively.
 The grayed region is the CEL used for the comparison CMD in the SFH analysis.
 The stripped region in the last 300~Myr shows the metallicity range that
 gave the lowest $\chi^2$ when determining the SFH. 
 The right-hand scale was converted from the abundance by mass assuming
 $Z_\Sun=$~0.0198.
\label{CEL}}
\end{figure}

    In short, the SFH is derived from the comparison of the star distribution
    of the observed CMD with that of a synthetic CMD.
    The synthetic CMD to be used in the comparison needs to cover a range of
    age and metallicity at least as large as the one that can be expected in
    such a dwarf irregular galaxy. It was generated by IAC-STAR \citep{apa04}
    using the stellar evolution library of \citet{ber94} and the
    bolometric corrections from \citet{cas03}, with the following
    parameters: the SFR was chosen constant between 13~Gyr ago and now, the
    initial mass function (IMF) was that of \citet{kro93}, and the
    fraction of binaries was set to zero. To fix the input CEL we
    fitted isochrones of different metallicities to the upper RGB of the
    outer field where crowding is not dominant.
    We used the \citet{gir02} isochrones for different ages (1.585,
    2.239, 5.012, 10.00 and 12.59~Gyr old) and metallicities between $Z=0.0005$
    and $Z=0.006$ in steps of 0.0005. The metallicities that are not available
    in the original library were interpolated using IAC-STAR.
    A metallicity was considered valid at a given age if the
    corresponding isochrone was inside the FWHM of the RGB between $20.4 \le
    I \le 21$. Figure~\ref{CEL} ({\it filled diamonds}) shows the values
    obtained through this method.
    Although the method is rather simplistic, our resulting CEL is
    in very good agreement with those derived by \citetalias{ski03}.
    However, \citet{gal05} showed that the theoretical isochrones are generally
    too vertical with respect to the empirical ones, leading to a slight
    overestimation of the metallicity. The grayed region represents the
    metallicity range employed to create the comparison CMD.
    It is mainly used for producing the RGB, from which we obtain the mean SFR
    between 1500 and 13000~Myr, and serves a normalization purpose for the young
    SFH.
    For the stars younger than 300~Myr, we tried to put constraints on their 
    metallicity by further restricting the metallicity range in the comparison
    CMD to intervals of 0.002 in abundance by mass, and ran the algorithm three
    times with the following ranges: $Z=$~0.0015-0.0035, $Z=$~0.0025-0.0045
    and $Z=$~0.0035-0.0055.
    
    In order to simulate the observational effects and allow a more realistic
    comparison with the real CMD, we applied the dispersion in color and
    magnitude of the synthetic stars recovered from the crowding tests to the
    comparison CMD, following the same procedure as in \citet{gal99}.
 
    In the observed and comparison CMDs, the MS, BL and RSG sequence are
    then divided into `boxes'. Because the box selection could influence the
    resulting SFH, we used three different types of box to rule
    out this possibility: two regular grids with large
    (0.3 and 0.5 in color and magnitude, respectively) and small (0.2 and 0.3)
    box size, and an ``\`a la carte'' parametrization. In the latter, the
    size and shape of the boxes are chosen taking into account
    the knowledge and model limitations of the evolutionary phases, in
    particular the slope of the RSG sequence and the position of the BL. The
    regular grids were also shifted in color and magnitude to check for
    consistency and ensure the significance of the result.
    
    Additionally, the synthetic stars are divided into partial models,
    each with a small range in age.
    The temporal resolution is limited by the quantity of information contained
    in the CMD, which depends on the quality of the data. It was chosen by
    comparing the capacity of the algorithm to recover the known SFH of a
    synthetic CMD, after the observational effects had been simulated as
    described above, using different time ranges. 
    Above a certain time resolution---which depends on age---the recovered
    SFH was mainly made of short, violent bursts of star formation separated
    by periods of inactivity, regardless of the input SFH \citep[see 
    also][]{apa07}.

    The reconstructed SFH is a linear combination of the different partial
    models. The best solution is obtained through $\chi^2$-minimization by
    a genetic code. A thorough description of the algorithm and method is
    presented in \citet{apa07}. 
    In total, we used 24 models with different CMD parametrization and
    time resolution. The consistency of the method was checked by
    solving the SFH of synthetic CMDs and comparing the solutions with the
    input SFHs.
    Some of the regular grid models did not give a satisfactory solution, but
    the discrepancies could generally be traced to small differences in the
    location of evolutionary phases in the CMD, while the solutions obtained
    with the {\it \`a la carte} parametrization were found to be more stable
    and are the ones we will present here.

\subsection{The Results}

\begin{figure*}
\epsscale{0.75} 
\epsscale{0.80} 
\plotone{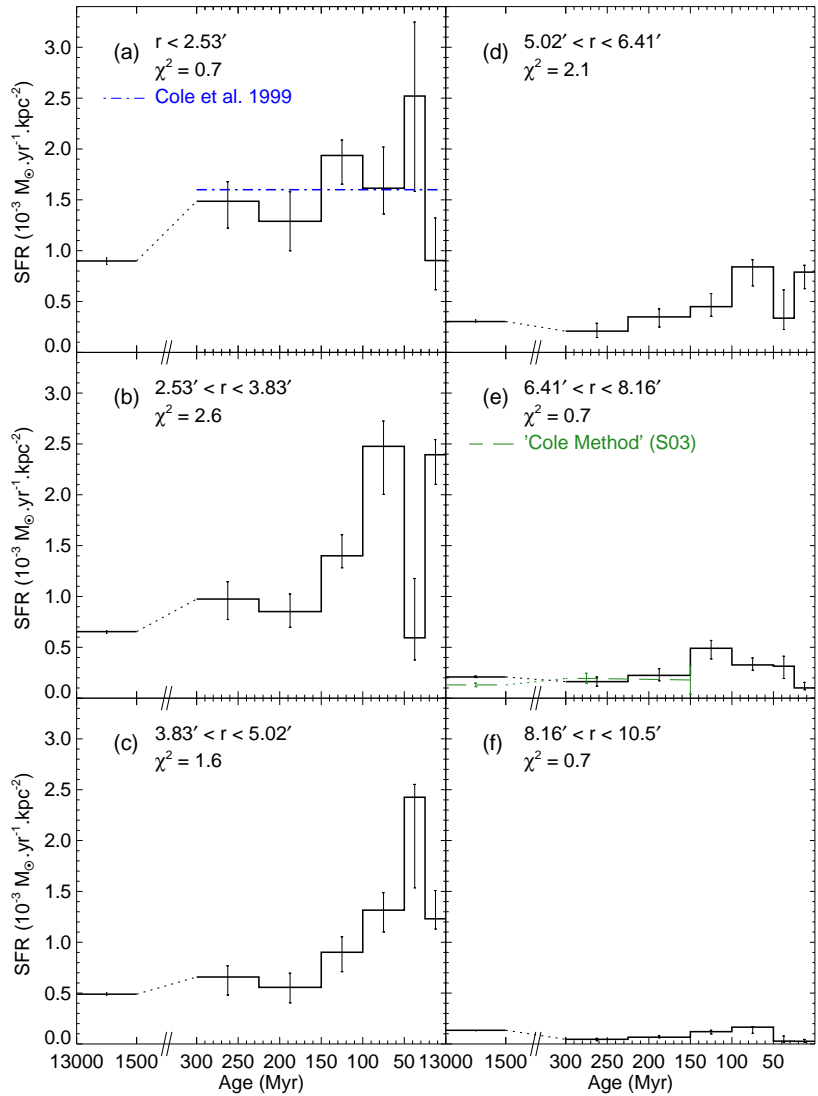}
\figcaption{Best SFR(t) obtained for each galactocentric distance.
 The radial ranges were chosen so that the corresponding CMDs contain the 
 same number of stars. The error bars correspond to the dispersion of twenty
 solutions for which $\chi^2_\sigma$ = $\chi^2+1$, where $\chi^2$ is the
 residual of the best solution, indicated in each panel. The gap in the 
 solutions between 300 and $\sim$$1500$~Myr is due to the lack of
 information from either the young populations or the RGB in this age range.
 The solutions obtained by \citet{col99} and \citet{ski03} for their 
 respective WFPC2 field are plotted in panels {\it (a)} and {\it (e)}.
\label{SFH}}
\end{figure*}

    Of the three metallicity ranges used for the stars younger than 300~Myr,
    the second one, i.e., $Z=$~0.0025-0.0045, gave the best solutions, and is
    shown as a stripped region in Fig.~\ref{CEL}. This is consistent with the
    H\,\textsc{II} region metallicity of \citet{tal80}.
    Our best solutions for the SFHs at different galactocentric distances are
    presented in Fig.~\ref{SFH}. The radius ranges were chosen so that the CMDs
    contain the same number of stars.
    The error bars correspond to the dispersion of twenty solutions for which
    $\chi^2$ = $\chi^2_{best}+1$ \citep[see][]{apa07}.

    The overall picture is a relatively constant SFR at all radii, decreasing
    with increasing radius, while the mean age of the stars increases with
    radius.
    The very central region, panel {\it (a)}, shows a fairly constant SFR
    for the last 300~Myr of 
    $(1.6 \pm 0.8) \times 10^{-3}  $~M$_\Sun~$yr$^{-1}~$kpc$^{-2}$,
    in excellent agreement with the value found by \citet[$1.6\times 10^{-3}
    $~M$_\Sun~$yr$^{-1}~$kpc$^{-2}$;][]{col99} using the $V$-band luminosity
    function of their WFPC2 central field and assuming a Salpeter IMF. The
    sharp drop in the last $\sim$25~Myr is in agreement with the lack of very
    bright stars at the center of IC1613 first noted by \citet{hod91}.
    
    Between $2\farcm5$ and $\sim$6$\arcmin$ [{\it (b)-(d)}] the SFR
    increased by a factor $\sim$2-3 in the last 100-150~Myr. This corresponds
    to the peak in the radial distribution of young stars and H\,\textsc{II}
    regions found by \citet{hod90}.

    The field studied by \citetalias{ski03} is a small fraction
    of the region for which the SFH is represented in panel {\it (e)}.
    The SFH calculated via the `Cole method'\footnote[7]{Three methods were
    used in \citetalias{ski03} to calculate the SFH: the `Dolphin method'
    gives the SFR on a relative scale only. 
    For the `Tolstoy method', the vertical scale of their Fig.~7 gives a mean
    SFR~$\sim3\times 10^{-3}$~M$_\Sun~$yr$^{-1}~$kpc$^{-2}$, a factor
    $\sim$15--20 higher than that of the `Cole method' and as high as the mean
    value of the SFR over the whole galaxy \citep{mat98}.}
    in \citetalias{ski03}, shown as the
    green long-dashed lines, is very similar to the one we obtained here for
    the whole elliptical annulus. 
    At larger radii, the number of stars formed more recently than about
    300~Myr decreases to a negligible value when the radius reaches
    $\sim$10$\arcmin$.

\section{Conclusions}\label{concl}

   We have presented an analysis of the stellar content, morphology,
   and recent star formation history of the Local Group dIrr galaxy IC1613
   based on wide-field ($V$-$I$,$I$) photometry of resolved stars.

   The distribution of resolved stars can be fitted with ellipses of
   position angle 80$\degr$ and eccentricity 0.15. The exponential fit of the
   resulting radial profile has a scale length of $2\farcm9 \pm 0.1$.
   The relatively large number of RGB stars still present in the outer part of
   our observed field ($r > 16\farcm5$) indicates that the galaxy is actually
   more extended than previously estimated.

   The well-populated young evolutionary phases of the CMD of the central
   region of IC1613 indicate very recent star formation ($\la$10~Myr).
   The changing CMDs as a function of galactocentric distance show a strong
   gradient in the age of the younger stellar population, with the young stars
   lying preferentially in the central part, while the older ones are distributed
   more uniformly. No evidence of star formation more recent than about
   300~Myr was detected beyond $r \ga 10\arcmin$. 
   
   Analysis of the width of the RGB as a function of radius is consistent
   with no metallicity gradient. The combination of crowding effect and the
   presence of younger stars in the RGB is responsible for the widening toward
   the center of the galaxy.

   In the region where the recent SFH could be studied ($r \la 10\arcmin$),
   the results indicate a decreasing SFR(t) from the center outward as
   expected from the distribution of neutral gas in IC1613, with the
   exception of the annular region where the star forming regions are clustered
   ($r\sim4\arcmin$) and therefore the SFR is a factor $\sim$2-3 higher.

\acknowledgments

 We would like to thank Dr.~A. Udalski and the OGLE collaboration for sharing
 their $VI$ photometry of IC1613, and the anonymous referee for valuable
 comments. 
 This research project has been supported by a Marie Curie Early Stage Research
 Training Fellowship of the European Community's Sixth Framework Programme under
 contract number MEST-CT-2004-504604, the IAC (grant P3/94) and the Spanish
 Education and Science Ministry (grant AYA2004-06343).
 As the reviewing of this article was near an end, our friend and coauthor
 Maurizio passed away in a tragic accident. His ideas and personality will
 be missed by many. 

\end{document}